\documentclass[aps,prl,twocolumn,superscriptaddress,showpacs,floatfix,nobibnotes]{revtex4}
\usepackage{epsfig}
\usepackage{epstopdf}

\usepackage{graphicx}
\usepackage{longtable}
\usepackage{CJK}
\usepackage{color}

\usepackage{mathptmx, courier, pifont}
\usepackage[scaled=0.92]{helvet}
\usepackage[T1]{fontenc}
\usepackage{textcomp}

\begin{document}

\begin{CJK*}{GB}{}

\title{The emergent dynamical symmetry at the triple point of nuclear deformations}

\author{Yu Zhang }
\affiliation{Department of Physics, Liaoning Normal University,
Dalian 116029, P. R. China}

\author{Feng Pan}
\affiliation{Department of Physics, Liaoning Normal University,
Dalian 116029, P. R. China}\affiliation{Department of Physics and
Astronomy, Louisiana State University, Baton Rouge, LA 70803-4001,
USA}

\author{Yu-xin Liu }
\affiliation{Department of Physics and State Key Laboratory of
Nuclear Physics and Technology, Peking University, Beijing 100871,
P. R. China} \affiliation{Collaborative Innovation Center of Quantum
Matter, Beijing 100871, China} \affiliation{Center for High Energy
Physics, Peking University, Beijing 100871, China}

\author{Yan-An Luo }
\affiliation{School of Physics, Nankai University, Tianjin 300071,
P. R. China}

\author{J. P. Draayer}
\affiliation{Department of Physics and Astronomy, Louisiana State
University, Baton Rouge, LA 70803-4001, USA}
\date{\today}

\date{\today}

\begin{abstract}

Based on the boson realization of the Euclidean algebras, it is
shown that the five-dimensional Euclidean dynamical symmetry may
emerge at the triple point of the shape phase diagram of the
interacting boson model, which thus offers a symmetry-based
understanding of this isolated point. It is further shown that the
low-lying dynamics in $^{108}$Pd, $^{134}$Ba, $^{64}$Zn, and
$^{114}$Cd may be dominated by the Euclidean dynamical symmetry.
\end{abstract}

\maketitle

\end{CJK*}

\begin{center}
\vskip.2cm\textbf{I. INTRODUCTION}
\end{center}\vskip.2cm

Dynamical symmetries (DSs) provide considerable insight into the
nature of quantum many-body dynamical structures. Generally, DS
occurs when the Hamiltonian of a system can be written in terms of
Casimir operators of a chain of Lie algebras $G\supset
G^\prime\supset G^{\prime\prime}\cdots$. Typical examples of DS are
those associated with the interacting boson model
(IBM)~\cite{IachelloBook87} for nuclear structure and the vibron
model (VM)~\cite{IachelloBook95} for molecules and atomic
clusters~\cite{HJP2006}.

The IBM possesses an overall U(6) symmetry with three DSs
corresponding to three typical collective structures or quadrupole
deformations~\cite{IachelloBook95}; namely, a spherical vibrator
[U(5)], an axially symmetric prolate rotor [SU(3)], and a
$\gamma$-soft rotor [O(6)]. In addition, an axially symmetric oblate
rotor [$\mathrm{\overline{SU(3)}}$] can be involved in the IBM
dynamics if adopting an alternative SU(3) quandrupole operator in
contrast to that often used for the prolate rotor~\cite{Jolie2001}.
Besides these exact DSs, the partial dynamical symmetries (PDS)
~\cite{Leviatan1996,Leviatan2007,Ramos2009} and quasidynamical
symmetries (QDS)~\cite{Rowe2004,Rowe2004II,Rowe2004III} have also
been found to occur in the IBM. Indeed, it was found that the SU(3)
QDS~\cite{Bonatsos2010,Bonatsos2011} may emerge along the trajectory
in the IBM parameter space close to the Alhassid-Whelan arc of
regularity~\cite{Alhassid1991}, which has been empirically
confirmed~\cite{Jolie2004}. A link between PDS and QDS has also been
established recently via the method of quantum number
fluctuation~\cite{Kremer2014}. As there is a link between each DS in
the IBM and the quantum (shape) phase or quadrupole
deformation~\cite{IachelloBook87}, the (shape) phase transitions in
nuclei may be characterized as the quantum phase transitions (QPTs)
in between the different DSs in the
IBM~\cite{CJC2010,CJ2009,Casten2007}. Particularly, the QPT from
SU(3) to $\mathrm{\overline{SU(3)}}$ may exactly occur at the point
of O(6) DS~\cite{Jolie2001}. In addition, an isolate triple point,
at which three kinds of quadrupole deformations including the
spherical, prolate, and oblate shapes may coexist at the same
time~\cite{Jolie2002,Warner2002}, also emerges in the critical
region of the IBM. On the other hand, an algebraic model of the
Euclidean dynamical symmetry in 5-dimension (Eu(5)
DS)~\cite{ZLPSD2014} has recently been suggested to describe the
nuclei in the critical region. Especially, it was
shown~\cite{ZLPSD2014} that the results obtained from the simplest
version of the Eu(5) DS (unprojected) in the large-$N$ limit are the
same as those from the E(5) critical point symmetry (E(5) CPS) built
from the Bohr Hamiltonian with an infinite well
potential~\cite{Iachello2000}. However, the relation between the IBM
and the Eu(5) DS still remains to be revealed. In this work, we will
present an extensive analysis of the Eu(5) DS and clarify the
relation between the Eu(5) DS and the IBM.

\begin{center}
\vskip.2cm\textbf{II. THE BOSON REALIZATION OF \\THE EUCLIDEAN
ALGEBRA}
\end{center}\vskip.2cm

A Hamiltonian in the IBM framework is constructed from two kinds of
boson operators; namely, a $s$-boson with $J^\pi=0^+$ and a
$d$-boson with $J^\pi=2^+$~\cite{IachelloBook87}. The three DSs in
the IBM are characterized by three different chains of the U(6)
group~\cite{IachelloBook87}:
\begin{eqnarray}\nonumber
&&\mathrm{U(6)} \supset \mathrm{U(5)} \supset \mathrm{SO(5)} \supset \mathrm{SO(3)}\, ,\\
&&\mathrm{U(6)} \supset \mathrm{O(6)} \supset \mathrm{SO(5)} \supset
\mathrm{SO(3)}\, ,\\ \nonumber&&\label{SU3} \mathrm{U(6)} \supset
\mathrm{SU(3)} \supset \mathrm{SO(3)}\, .
\end{eqnarray} Then, the Hamiltonian with an explicit DS in the IBM can be written
in terms of the Casimir operators of the corresponding group chain. On
the other hand, it was shown \cite{ZLPSD2014} that one can use
the $d$-boson operator to construct the Casimir operator of the
Eu(5) group as
\begin{eqnarray}
&&\label{C2}\hat{C}_2[\mathrm{Eu}(5)]=\hat{n}_{d} + \frac{5}{2} -
\frac{1}{2}\left(\hat{P}^{\dag}_{d} + \hat{P}_{d} \right)\, ,
\end{eqnarray}
where $\hat{n}_{d}=\sum_{u} d^{\dag} _{u} d_{u}$, and $\hat{P}_{d} =
\sum_{u} (-)^{u} d_{u} d_{-u}$. Accordingly, the $d$-boson operator
can be also used to construct the fifteen generators of the Eu(5)
Lie algebra as
\begin{eqnarray}\nonumber\label{gEu(5)}
&&\hat{Q}_u^{(2)}=\frac{1}{\sqrt{2}}[\tilde{d}_u-d_u^\dag],\\
&&\hat{T}_u^{(\lambda)} =
\sqrt{2}(d^\dag\tilde{d})_u^{(\lambda)},~\lambda=1,3,
\end{eqnarray}
where $\tilde{d}_u=(-1)^ud_u$. It is evident that the Eu(5) algebra
is non-compact as seen from (\ref{gEu(5)}). It should be mentioned
that a geometric realization of the Eu(5) algebra in the collective
model may be constructed by using the quadrupole coordinates $q_u$
and the conjugate momenta $\tilde{p}_u$~\cite{Caprio2007}. The boson
algebraic and the collective geometric realizations may be linked by
implementing the $d$-boson operator with
$\tilde{d}_{u}=\frac{1}{\sqrt{2}}[q_u+i\tilde{p}_u]$ and
$d^\dag_{u}=\frac{1}{\sqrt{2}}[q_u-i\tilde{p}_u]$~\cite{ZLPSD2014}.

It can be proven that the above Eu(5) generators  satisfy the
commutation relations
\begin{eqnarray}\nonumber
&&[\hat{Q}_u^{(2)},\hat{Q}_v^{(2)}]=0,\\ \nonumber
&&[\hat{T}_u^{(\lambda)},\hat{Q}_v^{(2)}]=-\sqrt{\frac{4\lambda+2}{5}}\langle\lambda
u 2v|2u+v\rangle \hat{Q}_{u+v}^{(2)},\\ \nonumber
&&[\hat{T}_u^{(\lambda)},\hat{T}_{u^\prime}^{(\lambda^\prime)}]=-\sqrt{8(2\lambda+1)(2\lambda^\prime+1)}\sum_{k=\mathrm{odd}}
\left\{\begin{array}{cc} \lambda,\lambda^\prime,k \\
2,2,2\,\end{array}\right\}\times\, \\
&&~~~~~~~~~~~~~~~~~~~~~~~~~\langle\lambda u\lambda^\prime
u^\prime|k u+u^\prime\rangle~\hat{T}_{u+u^\prime}^{(k)}\, ,
\end{eqnarray}
and \begin{equation}
[\hat{Q}_u^{(2)},~\hat{C}_2[\mathrm{Eu}(5)]]=[\hat{T}_u^{(\lambda)},~\hat{C}_2[\mathrm{Eu}(5)]]=0\,
.
\end{equation}
One can further prove that the operators $\{\hat{T}^{(\lambda)}_{u}\}$ with
$\lambda=1,~3$ generate the SO(5) algebra, in which the angular
momentum operators defined by $\{\hat{L}_{u}=\sqrt{5}\hat{T}^{(1)}_{u}\}$
generate the SO(3) algebra. The Eu(5) algebra
may be characterized by the algebraic chain~\cite{Caprio2007}
\begin{equation}\label{GEu(5)}
\mathrm{Eu(5)}\supset \mathrm{SO(5)}\supset \mathrm{SO(3)}\, .
\end{equation}
In addition, $\{\hat{Q}_u^{(2)}\}$ generate the Abelian group
$\mathrm{T}_5$ of translations in the five dimensional space.
It is thus realized that the Eu(5) algebra is equivalent to
the semidirect sum of $\mathrm{T_5}$ and
$\mathrm{SO(5)}$, namely, $\mathrm{Eu(5)}=\mathrm{T_5}\oplus_s \mathrm{SO(5)}$
~\cite{Wybourne1974,Feinsilver1996,Bonatsos2008}. Accordingly,
there is another dynamical symmetry related with
the algebraic chain
\begin{equation} \mathrm{Eu(5)}\supset
\mathrm{T_5}\oplus_s\mathrm{SO(3)}\supset \mathrm{SO(3)}\, ,
\end{equation} in which the semidirect sum $\mathrm{T_5}\oplus_s\mathrm{SO(3)}$
is often used to denote the dynamical symmetry of a quadrupole-deformed rigid body~\cite{Ui1970}.
In the following, only the dynamical situation characterized by the algebraic chain (\ref{GEu(5)})
will be studied.

\begin{center}
\vskip.2cm\textbf{III. THE SU(1,1) EXPRESSION OF THE EU(5) DS AND
ITS LINK WITH THE E(5) CPS}
\end{center}\vskip.2cm

As analyzed in \cite{ZLPSD2014}, the spectral structure of
(\ref{C2}) in the large-$N$ limit coincides with
that generated from the E(5) CPS~\cite{Iachello2000}. Actually, the
form of the Eu(5) Casimir operator shown in (2) can be directly
translated from the Hamiltonian of the E(5) CPS via the SU(1,1)
reformulation. Specifically, the SU(1,1) Lie algebra generated by
$\hat{S}_\nu$, $\nu=0,~\pm$, satisfy the commutation
relations~\cite{Rowe2004IV,Rowe2005,Rowe2005II,Rowe2009}
\begin{equation}
[\hat{S}_-,\hat{S}_+]=2\hat{S}_0,~~~~[\hat{S}_0,\hat{S}_\pm]=\pm\hat{S}_\pm\,
.
\end{equation}
The Casimir operator of SU(1,1) can be written as
\begin{equation}
\hat{C}_2[\mathrm{SU(1,1)}]=\hat{S}_0(\hat{S}_0-1)-\hat{S}_+\hat{S}_-\,
.
\end{equation}
Let ${|\lambda,k\rangle;~k=0,~1,~2,\cdots}$ be the basis vectors of
the irreducible representation $\lambda$ of SU(1,1), of which
the matrix representation is determined by~\cite{Rowe2004IV,Rowe2005,Rowe2005II,Rowe2009}
\begin{eqnarray}
&&\hat{S}_0|\lambda,k\rangle=\frac{1}{2}(\lambda+2k)|\lambda,k\rangle,\\
&&\hat{S}_+|\lambda,k\rangle=\sqrt{(\lambda+k)(k+1)}|\lambda,k+1\rangle,\\
&&\hat{S}_-|\lambda,k\rangle=\sqrt{(\lambda+k-1)k}|\lambda,k-1\rangle,\\
&&\hat{C}_2[\mathrm{SU(1,1)}]~|\lambda,k\rangle=\frac{\lambda}{2}(\frac{\lambda}{2}-1)~|\lambda,k\rangle\,
.
\end{eqnarray}

For the $d$-boson realization of the SU(1,1) ~\cite{Pan1998},

\begin{eqnarray}\label{su11d}
&&\hat{S}_+^d=\frac{1}{2}\hat{P}_d^+,~~~\hat{S}_-^d=\frac{1}{2}\hat{P}_d,
~~\hat{S}_0^d=\frac{1}{2}(\hat{n}_d+\frac{5}{2})\, ,\end{eqnarray}
from which the quantum numbers $\lambda$ and $k$ can be expressed by
the seniority quantum number $\tau$ of the SO(5) and the number  of
the $d$-bosons $n_{d}$ as $\lambda=\tau+{5\over{2}}$
and $k={1\over{2}}(n_{d}-\tau)$. Then, the Casimir operator
of Eu(5) defined in (\ref{C2}) can be expressed as
\begin{equation}\label{C22}
\hat{C}_2[\mathrm{Eu}(5)]=2\hat{S}_0^d-(\hat{S}_+^d+\hat{S}_-^d)\, .
\end{equation}
Thus, one may choose to solve the eigenvalue problem of
$\hat{C}_2[\mathrm{Eu}(5)]$ within the subspace spanned by
the SU(1,1) basis vectors $\{|\lambda,k\rangle\}$.
In addition, the Eu(5) with the Casimir operator
given in (\ref{C22}) can easily be extended to the n-dimensional
case. Specifically, the Casimir operator of the n-dimensional
Euclidean group (Eu(n)) with $n=2l+1$ may be written as
\begin{equation}\label{C23}
\hat{C}_2[\mathrm{Eu}(n)]=2\hat{S}_0^l-(\hat{S}_+^l+\hat{S}_-^l)\, ,
\end{equation}
of which the SU(1,1) algebra is generated by~\cite{Pan2002}
\begin{eqnarray}\label{su11l}\nonumber
&&\hat{S}_+^l=\frac{1}{2}\hat{P}_l^\dag,~~~\hat{S}_-=\frac{1}{2}\hat{P}_l\\
&&\hat{S}_0^l=\frac{1}{2}(\hat{n}_l+\frac{2l+1}{2})\,
,\end{eqnarray} where $\hat{n}_l=\sum_ml_m^\dag l_m$, and
$\hat{P}_l=\sum_m(-1)^ml_ml_{-m}$, in which  $l^\dag_{m}$ ($l_{m}$)
are the creation (annihilation) operators of the $l$-bosons.
It is obvious that the
SU(1,1) algebra given in (\ref{su11d}) is only a special case of
(\ref{su11l}) with $l=2$. Accordingly, one can also
construct the Eu(n) algebra with the $l$-boson operators as
\begin{eqnarray}\nonumber\label{gEu(n)}
&&\hat{Q}_u^{(l)}=\frac{1}{\sqrt{2}}[\tilde{l}_u-l_u^\dag],\\
&&\hat{T}_u^{(\lambda)} =
\sqrt{2}(l^\dag\tilde{l})_u^{(\lambda)},~\lambda=1,3,...,2l-1\, ,
\end{eqnarray}
of which the $l=2$ case just corresponds to the Eu(5) algebra given
in (\ref{gEu(5)}).

On the other hand, the collective Bohr Hamiltonian of the E(5)
CPS~\cite{Iachello2000} is written as
\begin{eqnarray}
H_{E(5)}=-\frac{\hbar^2}{2B}\Big\{\frac{1}{\beta^4}
\frac{\partial}{\partial\beta}\beta^4\frac{\partial}{\partial\beta}
+\frac{1}{\beta^2}(\frac{1}{\mathrm{sin}3\gamma}\frac{\partial}{\partial\gamma}
\mathrm{sin}3\gamma\frac{\partial}{\partial\gamma}   \nonumber\\
-\frac{1}{4}\sum_{k}\frac{{L^\prime}_k^2}{[\mathrm{sin}(\gamma-\frac{2}{3}n\pi)]^2})
\Big\} + V(\beta)
\end{eqnarray}
with \begin{equation}
V(\beta) =  \left\{ \begin{array}{cc} 0 \, , & ~~~~ \beta\leq \beta_{W} \, ,  \\
\infty \, , & ~~~~ \beta>\beta_{W} \, . \end{array} \right.
\end{equation}
By writing the eigenfuctions
\begin{equation}
\Phi(\beta,\gamma,\theta)=f(\beta)\Psi(\gamma,\theta)\, ,
\end{equation}
one can get the angular equation
\begin{eqnarray}
& & [-\frac{1}{\mathrm{sin}3\gamma}\frac{\partial}{\partial\gamma}
\mathrm{sin}3\gamma\frac{\partial}{\partial\gamma}
+\frac{1}{4}\sum_{k}\frac{{L^\prime}_k^2}{[\mathrm{sin}(\gamma-\frac{2k\pi}{3})]^2}]
\Psi(\gamma,\theta) \quad \nonumber  \\
& = & \Lambda\Psi(\gamma,\theta) ,
\end{eqnarray}
with $\Lambda =\tau(\tau+3)$ and the radial equation
\begin{equation}\label{Hr}
[-\frac{\hbar^2}{2B}(\frac{1}{\beta^4}\frac{\partial}{\partial\beta}\beta^4
\frac{\partial}{\partial\beta}-\frac{\Lambda}{\beta^2})
+V(\beta)]f(\beta)=E f(\beta)\, .
\end{equation}
According to the analysis shown in
~\cite{Rowe2004IV,Rowe2005,Rowe2005II,Rowe2009}, the SU(1,1)
algebra can be alternatively defined in terms of the differential operators as
\begin{equation}\label{a1}
\hat{S}_{\pm}^\beta=\frac{1}{4}[\frac{\triangle^2}{a^2}-\frac{\Lambda}{(a\beta)^2}+(a\beta)^2\mp(2\beta\frac{\partial}{\partial\beta}+5)]\,
,
\end{equation}
\begin{equation}\label{a2}
\hat{S}_{0}^\beta=\frac{1}{4}[-\frac{\triangle^2}{a^2}+\frac{\Lambda}{(a\beta)^2}+(a\beta)^2]\,
,
\end{equation}
with $\triangle^2=\frac{1}{\beta^4}\frac{\partial}
{\partial\beta}\beta^4\frac{\partial}{\partial\beta}$ and
$\frac{1}{a^2}=\frac{\hbar^2}{2B}$. Then, the Hamiltonian associated
with (\ref{Hr}) in the infinite well can be written as
\begin{equation}\label{Hs}
H=2\hat{S}_0^\beta-(\hat{S}_{+}^\beta+\hat{S}_{-}^\beta)\, ,
\end{equation}
which is the same form as that shown in (\ref{C22}). This indicates
that the Eu(5) DS can be directly translated from the E(5) CPS at
the Hamiltonian level. Thus, solving the differential equation
(\ref{Hr}) is approximately equivalent to diagonalizing the
Hamiltonian (\ref{Hs}) within the subspace spanned by the basis
vectors of SU(1,1)~\cite{Rowe2004IV,Rowe2005,Rowe2005II,Rowe2009}.
Though it is not easy to translate the boundary conditions of the
infinite well into those in the algebraic description, it can
approximately be realized in the diagonalization of (\ref{C22}) with
a cut-off in the Hilbert space~\cite{ZLPSD2014}.

\begin{center}
\vskip.2cm\textbf{IV. THE APPROXIMATE EU(5) DS AT THE TRIPLE POINT
IN THE IBM}
\end{center}\vskip.2cm

\begin{figure}
\begin{center}
\includegraphics[scale=0.30]{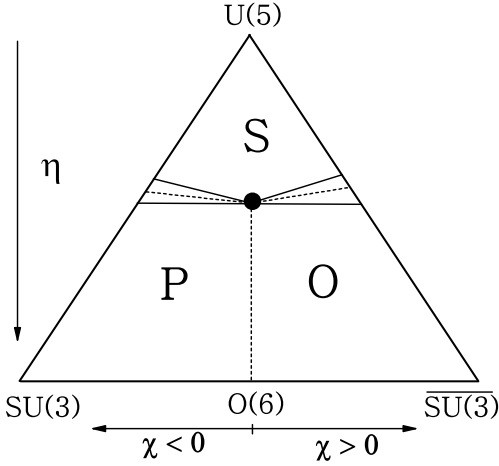}
\caption{(Color online) Shape phase diagram in the IBM parameter
space, where S represents the region with $\beta_\mathrm{e}=0$
corresponding to the spherical, P represents the region with
$\beta_\mathrm{e}>0$ corresponding to the prolate, and O represents
the region with $\beta_\mathrm{e}<0$ corresponding to the oblate. In
addition, the dashed lines correspond to the critical points of the
first-order QPTs, the two thin regions involving the dashed lines
represent the two-phase coexisting regions~\cite{Iachello1998}, and
the solid dot in the center represents the triple point. \label{F1}}
\end{center}
\end{figure}

Because the main composition of the system is the same as that of
the IBM with $d$-bosons at the second quantization level, to
investigate the relation between the IBM and the Eu(5) DS at the
Hamiltonian level, we consider the IBM consistent-$Q$
Hamiltonian~\cite{Warner1983}
\begin{equation} \label{Hamiltonian-IBM}
\hat{H}(\eta,~\chi)=\varepsilon \left[ (1-\eta)\hat{n}_{d} -
\frac{\eta}{4N}\hat{Q}^{\chi} \cdot \hat{Q}^{\chi} \right] \, ,
\end{equation}
where  $\hat{Q}^{\chi} = (d^{\dag} s + s^{\dag} \tilde{d})^{(2)} +
\chi (d^{\dag} \tilde{d})^{(2)}$ is the quadrupole operator, $\eta$
and $\chi$ are the control parameters with $\eta\in[0,1]$ and
$\chi\in[-\sqrt{7}/2,\sqrt{7}/2]$, and $\varepsilon$ is a scale
factor. It can be proven that the Hamiltonian is in the U(5) DS when
$\eta=0$; it is in the O(6) DS when $\eta=1$ and $\chi=0$; it is in
the SU(3) DS when $\eta=1$ and $\chi=-\frac{\sqrt{7}}{2}$;  and
it is in the $\overline{\mathrm{SU(3)}}$ DS when $\eta=1$ and
$\chi=\frac{\sqrt{7}}{2}$. The two-dimensional parameter space
of (\ref{Hamiltonian-IBM}) can be mapped onto a
symmetric triangle (see Fig.~\ref{F1}), called the extended Casten
triangle~\cite{Jolie2001}. To identify the QPTs in the IBM, one may
use the coherent state defined as~\cite{IachelloBook87}
\begin{eqnarray}
|\beta, \gamma, N\rangle&&=\frac{1}{\sqrt{N! (1 + \beta^2)^N}}
[s^\dag + \beta \mathrm{cos} \gamma d_0^\dag\, \nonumber \\&&  +
\frac{1}{\sqrt{2}} \beta \mathrm{sin} \gamma (d_2^\dag + d_{ -
2}^\dag)]^N |0\rangle\,
\end{eqnarray}
to obtain the scaled potential surface corresponding to the
Hamiltonian (\ref{Hamiltonian-IBM}) in the large-$N$ limit, which is
given as
\begin{eqnarray}\label{V}\nonumber
V_\mathrm{s}(\beta, \gamma)&&=\frac{1}{\varepsilon N}
\langle\beta, \gamma, N | H | \beta, \gamma, N\rangle|_{N\rightarrow\infty} \\
\nonumber &&= (1-\eta) \frac{\beta^2}{1+\beta^2} - \frac{\eta}{4(1 +
\beta^2)^2}\times\\ &&[4\beta^2 - 4\sqrt{\frac{2}{7}}\chi\beta^3
\mathrm{cos3} \gamma + \frac{2}{7}\chi^2\beta^4]\, .
\end{eqnarray}

To illustrate the type and the order of the QPTs, one should
minimize the potential function (\ref{V}) by varying $\beta$ and
$\gamma$ for $\eta$ and $\chi$. The optimal values are denoted as
$\beta_\mathrm{e}$ and $\gamma_\mathrm{e}$, from which one can get
the ground state energy per boson defined as
$E_g=V_s(\eta,\chi,\beta_\mathrm{e},\gamma_\mathrm{e})$. It can be
found that the $\gamma$-dependence in (\ref{V}) yields either
$\gamma_\mathrm{e}=0^\circ$ or $\gamma_\mathrm{e}=60^\circ$, of
which the case with $\gamma_\mathrm{e}=60^\circ$ can be equivalently
described by substituting $\gamma_\mathrm{e}=0^\circ$ and
$\beta=-\beta_\mathrm{e}$. Therefore, $\beta_e$ may serve as the
order parameter~\cite{Iachello2004} to identify the order and the
type of the QPTs. For the second-order QPT, the order parameter
$\beta_\mathrm{e}$ changes continuously, but with a discontinuous in
the second derivative of (\ref{V}). In contrast, the first-order QPT
may involve a discontinuous jump in the order parameter
$\beta_\mathrm{e}$ itself~\cite{Iachello2004}. In addition,
$\beta_\mathrm{e}=0$, $\beta_\mathrm{e}>0$ and $\beta_\mathrm{e}<0$
represent the spherical, prolate, and oblate deformations,
respectively. Based on the criteria mentioned above, one can prove
that the system may experience the first-order QPTs in two
directions with changing of the control parameters $\eta$ and
$\chi$~\cite{Jolie2002}. Specifically, the critical points of the
first-order QPTs occurring in the $\eta$ direction are given as
$\eta_c=\frac{14}{28+\chi^2}$ and $\chi\in[-\sqrt{7}/2,\sqrt{7}/2]$,
and in the $\chi$ direction are characterized by $\chi_c=0$ with
$\eta\in(0.5,1]$. Particularly, the crossing point of the
first-order QPTs occurring in the two directions, namely
$(\eta=0.5,\chi=0)$, may be recognized as the single triple point as
it is the junction point of the spherical, prolate, and oblate
deformations~\cite{Jolie2002}. Meanwhile, this point is also proven
to be the critical point of the second-order phase transition in the
$\eta$ direction. The whole shape phase diagram corresponding to
(\ref{V}) is clearly shown in Fig.~\ref{F1}.

 As is known, if a system has an underlying symmetry of the group G,
 the corresponding Hamiltonian
 should commutes with the generators of the group G.
 Along this line, it has been proven~\cite{Bonatsos2011} that there exists a parametrization
 trajectory preserving the approximate SU(3) symmetry (SU(3) QDS)
 inside the symmetry triangle of the IBM in the the large-$N$ limit.
 To identify the underlying Eu(5) DS in the IBM parameter space, we
 examine the commutation relations between the generators of the Eu(5) defined in (\ref{gEu(5)}) and the
 IBM Hamiltonian $\hat{H}(\eta,\chi)$ given in (\ref{Hamiltonian-IBM}).
 Firstly, it is easy to know that the IBM Hamiltonian does commute
 with the angular momentum operators $\hat{L}_u=\sqrt{5}\hat{T}_u^{(1)}$ since the Hamiltonian is a scalar.
 As a result, one only needs to examine the conditions under which the Hamiltonian
 may commute (approximately) with the other generators of the Eu(5).
 Specifically, one can prove the following commutation relations by using
 the standard angular momentum coupling techniques~\cite{Bonatsos2011,Edmondsbook}:
 \begin{eqnarray}\label{commutation}
 &&[\hat{T}_u^{(3)},\hat{n}_d]=0,\\
 &&[\hat{T}_u^{(3)},(d^\dag
 s+s^\dag\tilde{d})_v^{(2)}]\\ \nonumber
 &&=-\frac{\sqrt{14}}{5}\langle3u2v|2u+v\rangle(d^\dag
 s+s^\dag\tilde{d})_{u+v}^{(2)},\\
 &&[\hat{T}_u^{(3)},(d^\dag\tilde{d})_v^{(2)}]\\ \nonumber
 &&=2\sqrt{70}\sum_{k=2,4}\langle3u2v|ku+v\rangle
 \left\{ \begin{array}{cc}2~3~k  \\
 2~2~2 \end{array}\right\}(d^\dag\tilde{d})_{u+v}^{(k)},\\
 &&[\hat{Q}_u^{(2)},\hat{n}_d]=\frac{\sqrt{2}}{2}(\tilde{d}+d^\dag)_u^{(2)},\\
 &&[\hat{Q}_u^{(2)},(d^\dag
 s+s^\dag\tilde{d})_v^{(2)}]=(-)^u\delta_{u,-v}(s+s^\dag),\\
 &&[\hat{Q}_u^{(2)},(d^\dag\tilde{d})_v^{(2)}]=
 \frac{\sqrt{2}}{2}\langle2u2v|2u+v\rangle(\tilde{d}+d^\dag)_{u+v}^{(2)}\, .
 \end{eqnarray}
 By using the above relations, one can derive
 \begin{eqnarray}\label{CT3}
 &&[\hat{T}_q^{(3)},\hat{H}(\eta,\chi)]\\ \nonumber
 &&=\frac{3\sqrt{5}\varepsilon\eta\chi}{28N}
 \{\sqrt{10}[(\hat{B}^{(2)}\hat{A}^{(2)})_q^{(3)}-
 (\hat{A}^{(2)}\hat{B}^{(2)})_q^{(3)}]\\
 \nonumber
 &&-2[(\hat{B}^{(4)}\hat{A}^{(2)})_q^{(3)}-
 (\hat{A}^{(2)}\hat{B}^{(4)})_q^{(3)}]-
 2\chi[(\hat{B}^{(4)}\hat{B}^{(2)})_q^{(3)}\\
 \nonumber
 &&-(\hat{B}^{(2)}\hat{B}^{(4)})_q^{(3)}]\}\, ,
 \end{eqnarray}
where $\hat{A}_q^{(2)}=(s^\dag\tilde{d}+d^\dag s)_q^{(2)}$ and
$\hat{B}_q^{(k)}=(d^\dag\tilde{d})_q^{(k)}$ with $k=2,~4$.
Furthermore, by implementing the matrix elements related to the $s$-boson
operators under the
$\mathrm{U(6)}\supset \mathrm{U(5)}\supset \mathrm{SO(5)}\supset
\mathrm{SO(3)}$ basis vectors $\{|Nn_d\tau\Delta L\rangle\}$, where $N$,
$n_d$, $\tau$, and $L$ are the quantum number of U(6), U(5), SO(5),
and SO(3), respectively, and $\Delta$ is the additional quantum
number to characterize the multiplicity of $L$ in $\tau$, one gets
the replacements $s^\dag\rightarrow\sqrt{\hat{n}_s+1}$ and
$s\rightarrow\sqrt{\hat{n}_s}$ with $\hat{n}_s=N-\hat{n}_d$. Then,
 in the $n_d/N\ll1$ limit, one can derive that
\begin{eqnarray}\label{CQ2}\nonumber
&&[\hat{Q}_q^{(2)},\hat{H}(\eta,\chi)]=\frac{\sqrt{2}\varepsilon}{2}(1-2\eta)
\hat{C}_q^{(2)}-\\ \nonumber
&&\frac{\sqrt{2}\varepsilon\eta\chi}{8N}
[(\hat{A}^{(2)}\hat{C}^{(2)})_q^{(2)}+(\hat{C}^{(2)}\hat{A}^{(2)})_q^{(2)}+\\
&&2\hat{B}_q^{(2)}+\chi(\hat{C}^{(2)}\hat{B}^{(2)})_q^{(2)}+\chi(\hat{B}^{(2)}\hat{C}^{(2)})_q^{(2)}]\,
,
\end{eqnarray}
where $\hat{C}_q^{(2)}=(\tilde{d}+d^\dag)_q^{(2)}$. In order to make
the commutators given in (\ref{CT3}) and (\ref{CQ2}) vanish at the
same time, it is uniquely required $(\eta=0.5,\chi=0)$, under which
the IBM Hamiltonian just locates at the single triple point
mentioned above. The result clearly shows that the Hamiltonian at
the triple point is approximately invariant under the Eu(5)
transformations in the $n_d/N\ll1$ limit. It should be noted that
the approximation condition $n_d/N\ll1$ is well satisfied for
low-lying states generated from the IBM Hamiltonian
(\ref{Hamiltonian-IBM}) with $\eta\in[0,0.5]$ and $\chi=0$ in large
$N$ cases~\cite{Arias2003,Iachello2004,PZD2005}. In fact, if
implementing the matrix elements related to $s$-boson operators
under the basis vectors $\{|Nn_d\tau\Delta L\rangle\}$, one can
write the Hamiltonian (\ref{Hamiltonian-IBM}) at the triple point
($\eta=0.5,~\chi=0$) as
\begin{eqnarray} \label{Hamiltonian-tri}
&&\hat{H}_{\rm tri}=\frac{\varepsilon}{8}\{4\hat{n}_{d}-\frac{1}{N}
[\hat{n}_d(N-\hat{n}_d+1)+\nonumber\\
&&~~(N-\hat{n}_d)(\hat{n}_d+5)+d^\dag\cdot d^\dag\sqrt{(N-\hat{n}_d)(N-\hat{n}_d-1)}+\nonumber\\
&&~~\sqrt{(N-\hat{n}_d+1)(N-\hat{n}_d+2)}~\tilde{d}\cdot\tilde{d}]\}
\, .
\end{eqnarray}
In the $n_d/N\ll1$ limit, Eq.~(\ref{Hamiltonian-tri}) can be further
approximated as~\cite{Rowe2004}
\begin{eqnarray} \label{Hamiltonian-tri2}\nonumber
&&\hat{H}_{\rm tri}\simeq\frac{\varepsilon}{4}\left[\hat{n}_{d} -
\frac{5}{2} -\frac{1}{2}(P^{\dag}_{d} + P_{d} )\right]\\
&&~~~~~~=\frac{\varepsilon}{4}\left[\hat{C}_2[\mathrm{Eu}(5)]-5\right]\,
,
\end{eqnarray}
which is explicitly given as the Casimir operator of the Eu(5) up to
a constant and a scale factor. It is thus confirmed that the Eu(5)
DS indeed occurs in the $n_d/N\ll1$ limit at the triple point.
Rigorously speaking, the Eu(5) DS occurring at the triple point is
only almost exact for the ground state in the large-$N$ limit but
approximate for the excited states because the condition $n_d/N\ll1$
for the excited states at the triple point becomes weaker with the
increasing of the excitation energies. In addition, it should be
emphasized that the Eu(5) DS is the (approximate) DS associated with
a $\beta$-soft potential since the scale potential surface deduced
from (\ref{V}) at the triple point is soft in $\beta$ in contrast to
other DSs in the IBM, of which the potential surfaces are all
relatively rigid in $\beta$.

\begin{center}
\vskip.2cm\textbf{V. POSSIBLE EU(5) CANDIDATES}
\end{center}\vskip.2cm

\begin{figure*}
\begin{center}
\includegraphics[scale=0.5]{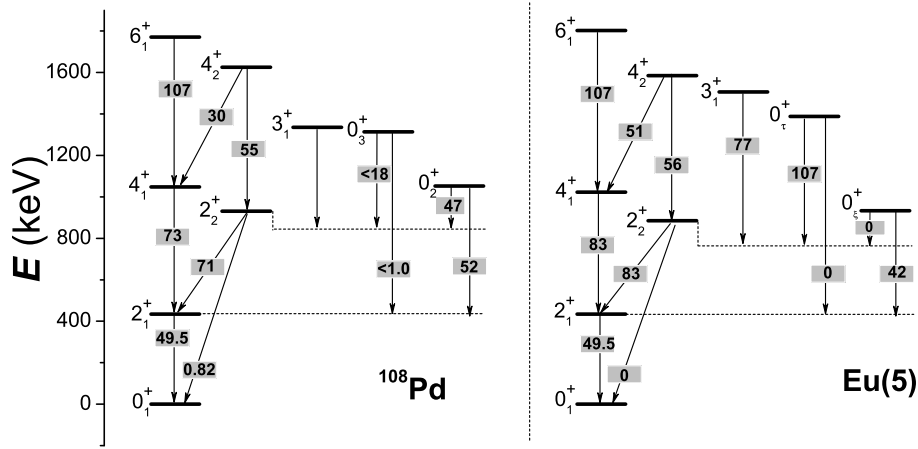}
\caption{(Color online) The low-lying structure of
$^{108}$Pd~\cite{Blachot1997} and the results calculated for the
Eu(5) Hamiltonian (\ref{HEu(5)}) with the $m=100$ truncation, where
$0_\tau^+$ and $0_\xi^+$ represent the excited $0^+$ state with
$\tau=3$ in the $\xi=1$ family and that with $\tau=0$ in the $\xi=2$
family, respectively, as those in the E(5) CPS~\cite{Iachello2000}.
The parameters involved in the Eu(5) Hamiltonian are set as
$\alpha=49.4$keV, $b=14.8$keV, and $c=9.9$keV.}\label{F2}
\end{center}
\end{figure*}

\begin{figure*}
\begin{center}
\includegraphics[scale=0.5]{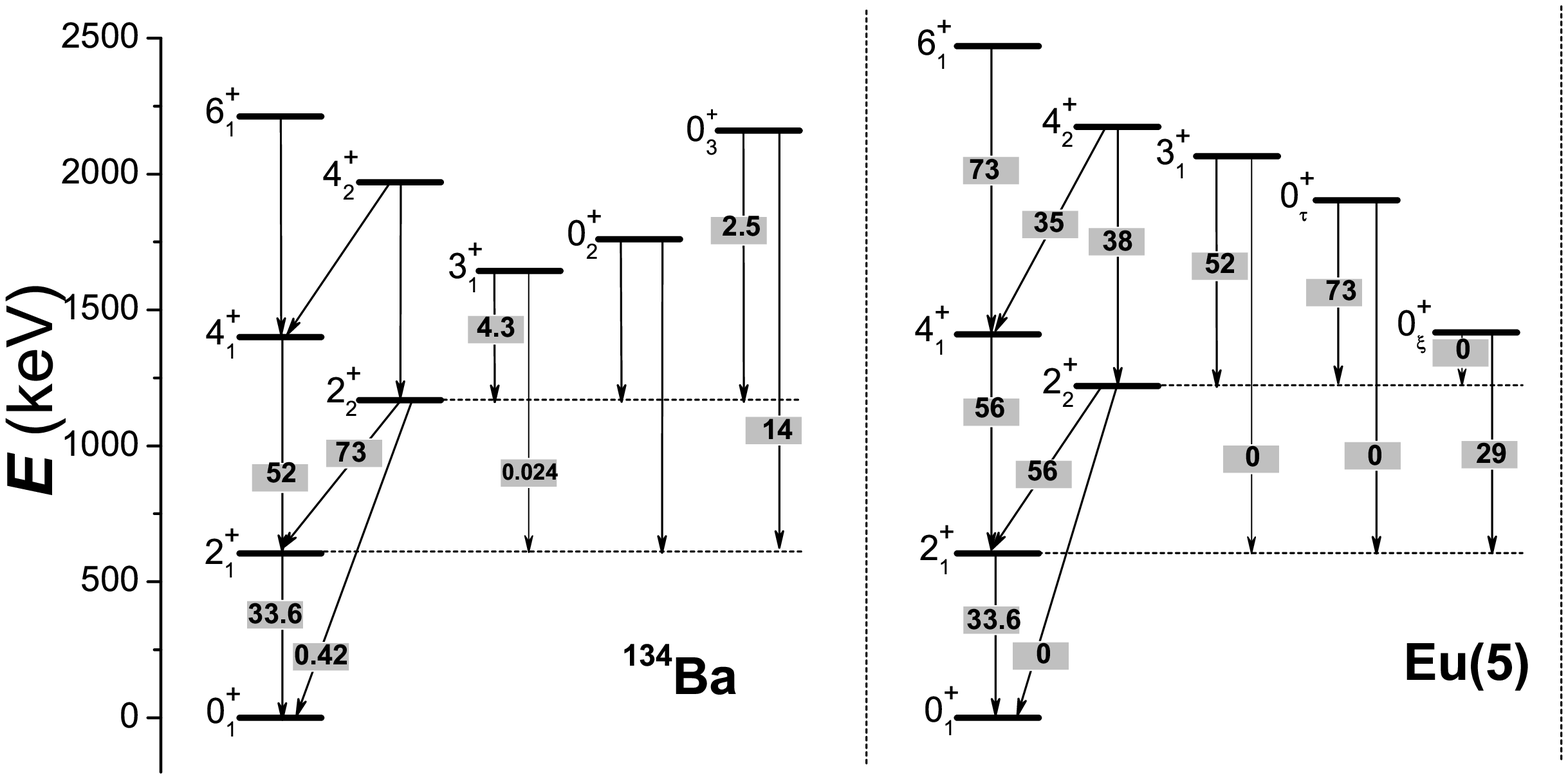}
\caption{(Color online) The same as in Fig.~\ref{F2} but for
$^{134}$Ba~\cite{Sonzogni2004} and the corresponding Eu(5) pattern,
for which the parameters are set as $\alpha=74.9$keV, $b=11.2$keV
and $c=13.5$keV.}\label{F3}
\end{center}
\end{figure*}

\begin{figure*}
\begin{center}
\includegraphics[scale=0.5]{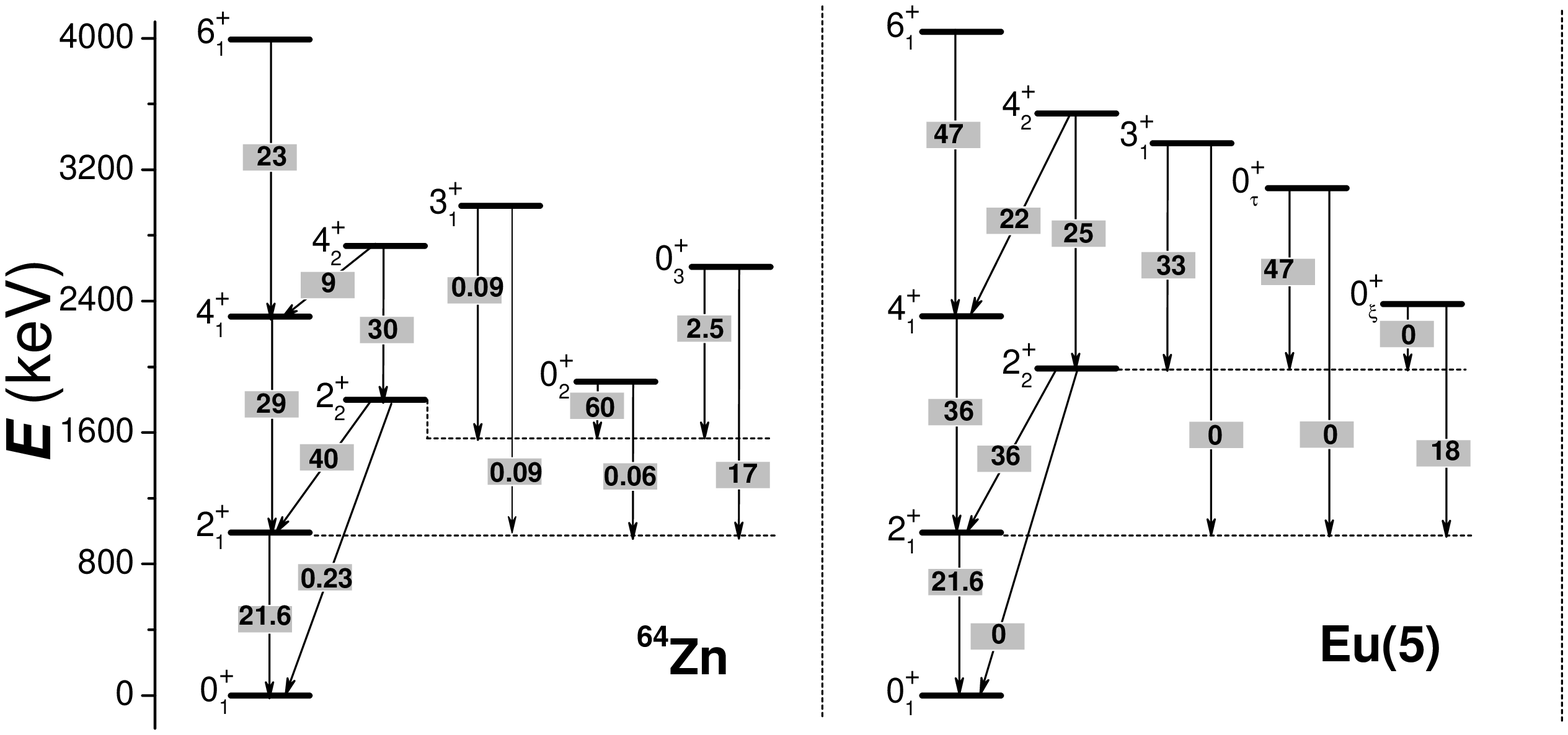}
\caption{(Color online) The same as in Fig.~\ref{F2} but for
$^{64}$Zn~\cite{singh1996} and the corresponding Eu(5) pattern, for
which the parameters are set as $\alpha=126$keV, $b=12.6$keV and
$c=22.7$keV.}\label{F4}
\end{center}
\end{figure*}

\begin{figure*}
\begin{center}
\includegraphics[scale=0.5]{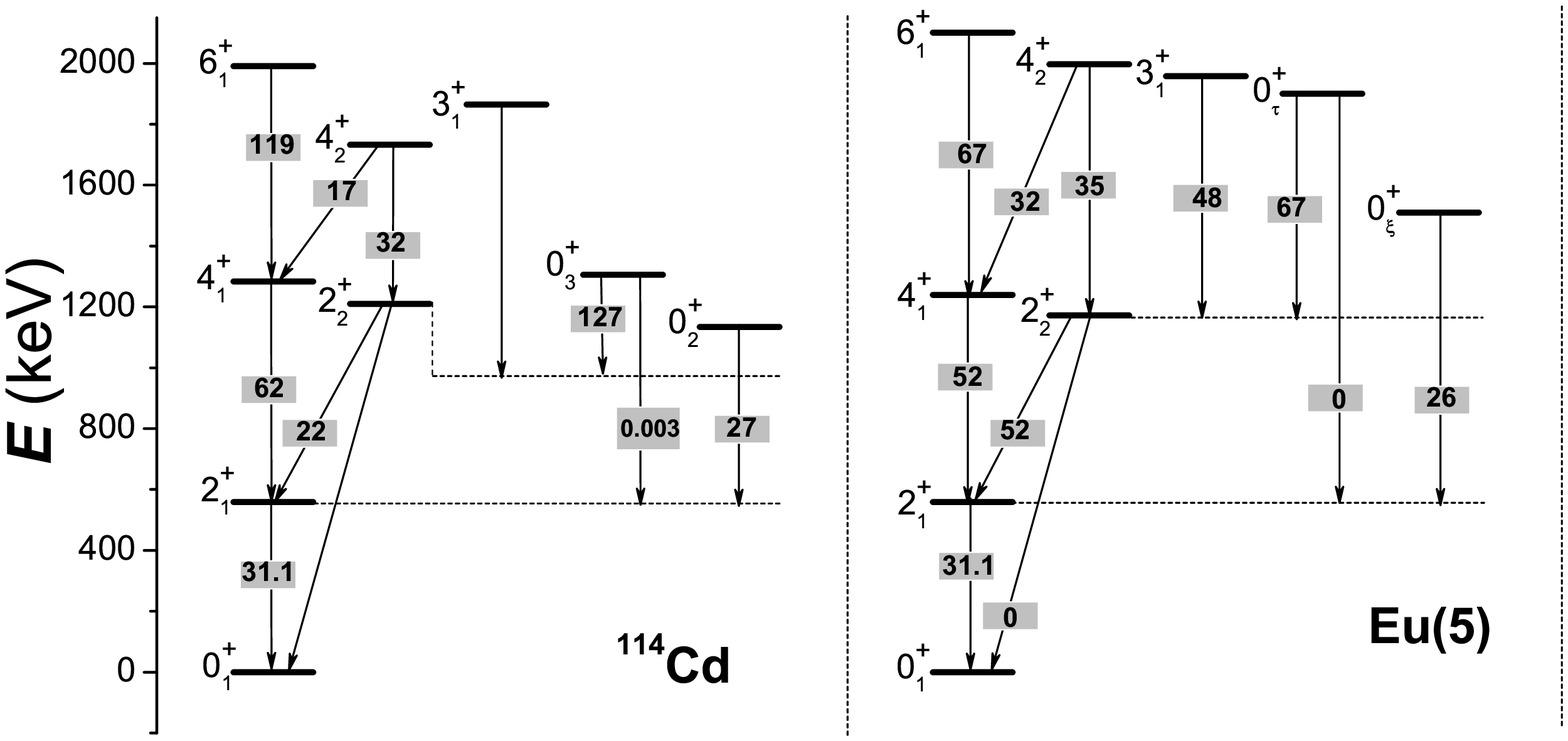}
\caption{(Color online) The same as in Fig.~\ref{F2} but for
$^{114}$Cd~\cite{Blachot2002} and the corresponding Eu(5) pattern,
for which the parameters are set as $\alpha=79.9$keV, $b=4.8$keV and
$c=4.8$keV.}\label{F5}
\end{center}
\end{figure*}

Based on the concept of dynamical symmetry, the Hamiltonian with the
Eu(5) DS can generally be written as
\begin{equation}\label{HEu(5)}\hat{H}_{\rm Eu(5)}=a~
\hat{C}_2[\mathrm{Eu}(5)]+b~\hat{C}_2[\mathrm{SO}(5)]+c~\hat{C}_2[\mathrm{SO}(3)]\,
,
\end{equation}
where $a$,~$b$, and $c$ are adjustable parameters, and
$\hat{C}_2[\mathrm{SO}(5)]$ and $\hat{C}_2[\mathrm{SO}(3)]$ are the
Casimir operators of SO(5) and SO(3) defined as
\begin{eqnarray}
&&\hat{C}_2[\mathrm{SO}(5)]=\hat{T}^3\cdot\hat{T}^3+\hat{T}^1\cdot\hat{T}^1,\\
&&\hat{C}_2[\mathrm{SO}(3)]=5\hat{T}^1\cdot\hat{T}^1\, ,
\end{eqnarray} with $\hat{T}^1$ and $\hat{T}^3$ being
those given in (\ref{gEu(5)}). One can construct
eigenstates of (\ref{HEu(5)}) from the SO(5) basis vectors
$\{|\tau \; \Delta L\rangle\equiv|n_{d}=\tau,\tau \; \Delta L\rangle\}
$, which is well defined in the IBM, since SO(5) is the
subalgebra of the Eu(5) as shown in (\ref{GEu(5)}).
Specifically, the eigenstates of the Hamiltonian in the
$d$-boson system with the seniority $\tau$ and
angular momentum $L$ being good quantum numbers
can be expressed as
\begin{equation}\label{WF}
|\xi \,\tau \; \Delta L\rangle=\sum_{k=0}^m~C_k^\xi
(\hat{P}_d^\dag)^k |\tau \; \Delta L\rangle\, ,
\end{equation}
where $m+1$ represents the dimension of the Hilbert subspace,
and $C_k^\xi$ is the expansion coefficient with $\xi$ being the
additional quantum number to distinguish the states
with the same $\tau$, $\Delta$, and $L$.
The expansion coefficients $\{C_k^\xi\}$
are determined by the eigen-equation
\begin{equation}
\hat{H}_{\rm Eu(5)}|\xi \,\tau \; \Delta L\rangle=E_{\tau~L}^\xi|\xi
\,\tau \; \Delta L\rangle\, . \end{equation} Concretely, one can
diagonalize the Hamiltonian (\ref{HEu(5)}) in the $m+1$ dimensional
Hilbert subspace to get the eigenvalues $E_{\tau~L}^\xi$ and the
corresponding  expansion coefficients $\{C_k^\xi\}$. Generally, for
the IBM Hamiltonian such as that of the O(6) DS,
$m=[\frac{N-\tau}{2}]$ is required, where $[x]$ is the integer part
of $x$~\cite{IachelloBook87}, and the $s$-boson part $| n_s\rangle$
should also be involved in  (\ref{WF}) for the IBM with
$n_s=N-2k-\tau$. In the present case, the $s$-boson part is
irrelevant, while $m$ should, in principle, be taken as infinite
since the dimension of the Hilbert subspace is infinite due to the
non-compactness of the Eu(5) algebra~\cite{Wybourne1974}. However,
the analysis \cite{ZLPSD2014} shows that the dynamical structure of
the Eu(5) DS may be well kept in the finite-$m$ cases, which
indicates that one can diagonalize the Eu(5) Hamiltonian within a
finite subspace  with sufficient large-$m$ truncation. In our
calculation, the parameter $a$ in (\ref{HEu(5)}) is reset as
$a=\alpha~m$ for convenience since the energy levels generated by
$\hat{C}_2[\mathrm{Eu}(5)]$ may scale with $m^{-1}$ as shown in
\cite{ZLPSD2014}. In contrast, the energy levels generated by
$H_{\mathrm{tri}}$ shown in (\ref{Hamiltonian-tri}) may scale with
$m^{-1/3}$~\cite{Dusuel2005}. Besides the diagonalization scheme, it
should be mentioned that the eigenstates corresponding to a DS in
the IBM may be built through the so called spectrum algebra method.
For example, the eigenstates in the O(6) DS can be analytically
constructed by acting the generalized boson pairing operators on the
O(6) basis vector~\cite{IachelloBook87}. Similarly, as shown in
(\ref{WF}), eigenstates of the Eu(5) DS are constructed by acting
the generalized boson pairing operator $\sum_{k=0}^m~C_k^\xi
(\hat{P}_d^\dag)^k$ on the SO(5) basis vectors with $m$ being
infinite. But the coefficients $C_k^\xi$ can be only calculated in a
numerical way at present.

In the previous work~\cite{ZLPSD2014}, it was shown that the
simplest version of Eu(5) DS involving only the first term in
(\ref{HEu(5)}) provides an algebraic description of the E(5)
CPS~\cite{Iachello2000}, in which the E(5) CPS has been widely
confirmed
~\cite{Casten2000,Frank2001,ZL2002,Zamfir2002,Clark2004,Mihai2007,Zhang2003,Conquard2009}.
It is thus suggested that the E(5) nuclei may be the candidate of
the Eu(5) DS in experiments. Here, we choose
$^{108}$Pd~\cite{Blachot1997}, $^{134}$Ba~\cite{Sonzogni2004},
$^{64}$Zn~\cite{singh1996} and $^{114}$Cd~\cite{Blachot2002}, which
were previously identified as the candidates of the E(5)
CPS~\cite{ZL2002,Casten2000,Mihai2007,Zhang2003}, as examples to
show the possible Eu(5) patterns in experiments.  In our
calculation, the E2 transition operator is taken as

\begin{equation}
\hat{T}_u=e(d^\dag+\tilde{d})_u^{(2)}
\end{equation}
with the effective boson charge $e$ determined by the corresponding
experiment value of $B(E2;2_1\rightarrow0_1)$ (in W.u.). The
low-lying patterns of these E(5) nuclei and the corresponding
results obtained from the Eu(5) DS are shown in
Fig.~\ref{F2}-\ref{F5}. It can be clearly observed from
Fig.~\ref{F2} that the low-lying spectrum of $^{108}$Pd can be well
described by the Eu(5) pattern determined by (\ref{HEu(5)}).
Particularly, the relative $B(E2)$ strengths in the Eu(5) DS are
independent of the parameters, while the data of $^{108}$Pd seem to
be well reproduced by those of the Eu(5) DS. In addition,
$B(E2;0_{3,2}^+\rightarrow2_2^+)$ evidently deviates from
$B(E2;0_\tau^+\rightarrow2_2^+)$ in the Eu(5) DS, which indicates
that the $0_\xi^+$ and $0_\tau^+$ components may be mixing in the
excited $0^+$ states of $^{108}$Pd. As shown in
Fig.~\ref{F3}-\ref{F5}, the low-lying pattern of $^{134}$Ba,
$^{64}$Zn, and $^{114}$Cd can also be globally reproduced well in
the Eu(5) DS, which further confirms that the Eu(5) DS emerges in
these nuclei. Meanwhile,  deviations from the experimental results
still exist. For example, $B(E2;3_1^+\rightarrow2_2^+)$ calculated
from the Eu(5) DS description seems too large, and the ordering of
the first two excited $0^+$ levels in $^{134}$Ba and $^{64}$Zn is
altered in the Eu(5) DS as show in Figs.~\ref{F3} and ~\ref{F4},
which show that the Eu(5) DS is still an approximate symmetry. As
analyzed in \cite{Bonatsos2006}, the order of the first two excited
$0^+$ states can be altered by using a $\gamma$-independent
displaced infinite well $\beta$ potential in the Bohr Hamiltonian in
contrast to the one used in the E(5) CPS. It would be very
interesting to investigate whether such an improvement in the E(5)
CPS will be achieved in the Eu(5) DS, which may be discussed
elsewhere.

In comparison of the Eu(5) DS results with those of the E(5) and the
IBM consistent-Q Hamiltonian at the triple point, some typical level
energies and $B(E2)$ ratios calculated from these models in
comparison with the corresponding experimental data are shown in
Table~\ref{T1}. As clearly shown in Table~\ref{T1}, the results of
these models are similar and accord with the experimental data with
quantitative differences.
In addition, one may notice from Table~\ref{T1} that the $B(E2)$
ratios obtained from the E(5) CPS are almost the same as those
calculated from the Eu(5) DS with $m=100$ up to the second decimal
place. Actually, the minor differences at the second decimal place
in the $B(E2)$ ratios shown in Table~\ref{T1} can also be removed if
these quantities are calculated with larger $m$ truncation, which
indicates that the Eu(5) Hamiltonian (\ref{HEu(5)}) is just an
algebraic equivalent description of the E(5) CPS with SO(5) and
SO(3) invariants being involved, with which an intimate relation
between the E(5) CPS and the IBM beyond the mean-field
approximation~\cite{AAVGDF2003,Ramos2008,Zhang2011} is thus
revealed. It should be mentioned that one can also eliminate the
differences of the energy ratios in the Eu(5) DS from the E(5)
description shown in Table~\ref{T1} by adding a linear combination
of  the SO(5) and SO(3) Casimir operators  in the E(5) CPS
Hamiltonian~\cite{Iachello2000,Caprio2007}.

\begin{table}[htb]
\caption{Typical energy and $B(E2)$ ratios of
$^{108}$Pd~\cite{Blachot1997}, $^{134}$Ba~\cite{Sonzogni2004},
$^{64}$Zn~\cite{singh1996}, and $^{114}$Cd~\cite{Blachot2002}
calculated from the Eu(5) DS extracted from Fig.~\ref{F2}-\ref{F5},
the IBM at the triple point with the boson number $N$ taken as the
number of valence nucleon (or hole) pairs for each nucleus, and the
E(5) CPS~\cite{Iachello2000} in comparison with the corresponding
experimental data. } \label{T1}
\begin{tabular}{cccccc}\hline\hline
&$\frac{E(4_1^+)}{E(2_1^+)}$&$\frac{E(2_2^+)}{E(4_1^+)}$&$\frac{E(0_\xi^+)}{E(2_1^+)}$
&$\frac{B(E2;4_1^+\rightarrow2_1^+)}{B(E2;2_1^+\rightarrow0_1^+)}$&$\frac{B(E2;0_2^+\rightarrow2_1^+)}{B(E2;2_1^+\rightarrow0_1^+)}$\\
\hline
$^{108}$Pd&2.42&0.89&2.43&1.48&1.05\\
Eu(5)&2.36&0.86&2.15&1.67&0.85\\
Tri($N=8$)&2.14&1.00&3.00&1.55&0.81\\
E(5)&2.20&1.00&3.03&1.68&0.86\\
\hline
$^{134}$Ba&2.32&0.83&3.57&1.55&0.42\\
Eu(5)&2.33&0.87&2.34&1.67&0.85\\
Tri($N=5$)&2.15&1.00&3.18&1.41&0.66\\
E(5)&2.20&1.00&3.03&1.68&0.86\\
\hline
$^{64}$Zn&2.33&0.78&2.63&1.34&0.79\\
Eu(5)&2.33&0.86&2.40&1.67&0.85\\
Tri($N=4$)&2.16&1.00&3.29&1.32&0.58\\
E(5)&2.20&1.00&3.03&1.68&0.86\\
\hline
$^{114}$Cd&2.30&0.94&2.03&1.99&0.87\\
Eu(5)&2.22&0.95&2.70&1.67&0.85\\
Tri($N=9$)&2.14&1.00&2.96&1.57&0.84\\
E(5)&2.20&1.00&3.03&1.68&0.86\\
\hline\hline
\end{tabular}
\end{table}

\vskip 1.3cm
\begin{center}
\vskip.2cm\textbf{VI. SUMMARY}
\end{center}\vskip.2cm

In summary, the boson realization of the Eu(5) algebras has been
presented. Based on which the relation between the Eu(5) dynamical
symmetry description and the IBM is discussed. Specifically, it is
shown that the Eu(5) dynamical symmetry may emerge at the triple
point of the IBM phase diagram in the $n_d/N\ll1$ limit, which thus
provides an alternative insight into the dynamical structure
at/around this isolated point and the experimental data associated
with it. On the other hand, this work also shows that the results of
the E(5) CPS  can be realized in a fully DS way within the algebraic
frame, which thus reveals a more intimate relationship between the
E(5) CPS and the IBM beyond the mean-field approximation. Finally, a
preliminary examination of the Eu(5) DS in $^{108}$Pd, $^{134}$Ba,
$^{64}$Zn, and $^{114}$Cd is made. The results indicate that the
low-lying dynamics of these nuclei are indeed dominated by the Eu(5)
DS.

\bigskip

\begin{acknowledgments}
Supports from U.S. National Science Foundation (OCI-0904874),
Southeastern Universities Research Association, the Natural Science
Foundation of China (11375005, 11005056, 11175078, 11175004,
11435001 and 11475091), and the LSU--LNNU joint research program
(9961) are acknowledged. YXL thanks also the support from the
National Key Basic Research Program of China under Contract No.
G2013CB834400.
\end{acknowledgments}




\begin{thebibliography}{99}


\bibitem{IachelloBook87}
        F. Iachello, and A. Arima, {\it The Interacting Boson Model}
        (Cambridge University, Cambridge, England, 1987).


\bibitem{IachelloBook95}
        F. Iachello, and R. D. Levine, {\it Algebraic Theory of Molecules}
        (Oxford University, Oxford, UK 1995).

\bibitem{HJP2006}
        H. Y\'{e}pez-Mart\'{i}nez, J. Cseh, and P. O. Hess,
        Phys. Rev. C {\bf 74}, 024319 (2006).

\bibitem{Jolie2001}
        J. Jolie, R. F. Casten, P. von Brentano, and V. Werner, Phys. Rev. Lett. {\bf 87}, 162501 (2001).

\bibitem{Leviatan1996}
        A. Leviatan, Phys. Rev. Lett. {\bf 77}, 818 (1996).

\bibitem{Leviatan2007}
        A. Leviatan, Phys. Rev. Lett. {\bf 98}, 242502 (2007).

\bibitem{Ramos2009}
        J. E. Garc\`{i}a-Ramos, A. Leviatan, and P. Van Isacker, Phys. Rev. Lett. {\bf 102}, 112502 (2009).


\bibitem{Rowe2004}
        D. J. Rowe, Phys. Rev. Lett. {\bf 93}, 122502 (2004).

\bibitem{Rowe2004II}
        D. J. Rowe, P. S. Turner, and G. Rosensteel, Phys. Rev. Lett.
        {\bf 93}, 232502 (2004).

\bibitem{Rowe2004III}
        D. J. Rowe, Nucl. Phys. A {\bf 745}, 47 (2004).

\bibitem{Bonatsos2010}
        D. Bonatsos, E. A. McCutchan, and R. F. Casten,
        Phys. Rev. Lett. {\bf 104}, 022502 (2010).

\bibitem{Bonatsos2011}
        D. Bonatsos, S. Karampagia, and R. F. Casten,
        Phys. Rev. C {\bf  83}, 054313 (2011).

\bibitem{Alhassid1991}
        Y. Alhassid and N. Whelan,
        Phys. Rev. Lett. {\bf 67}, 816 (1991).

\bibitem{Jolie2004}
        J. Jolie, {\it et al.},
        Phys. Rev. Lett. {\bf 93}, 132501 (2004).


\bibitem{Kremer2014}
        C. Kremer, {\it et al.},
        Phys. Rev. C {\bf 89}, 041302(R) (2014).

\bibitem{CJC2010}
        P. Cejnar, J. Jolie, and R. F. Casten, Rev. Mod. Phys. {\bf 82},
         2155 (2010).

\bibitem{CJ2009}
        P. Cejnar, and J. Jolie, Prog. Part. Nucl. Phys. {\bf 62},
         210 (2009).

\bibitem{Casten2007}
        R. F. Casten and E. A. McCutchan, J. Phys. G {\bf 34}, R285 (2007).

\bibitem{Jolie2002}
        J. Jolie, P. Cejnar, R. F. Casten, S. Heinze, A. Linnemann, and V. Werner, Phys. Rev. Lett. {\bf 89}, 182502 (2002).

\bibitem{Warner2002}
        D. Warner, Nature {\bf 42}, 614 (2002).

\bibitem{ZLPSD2014}
        Y. Zhang, Y. X. Liu, F. Pan, Y. Sun, and J. P. Draayer, Phys. Lett. B {\bf 732}, 55 (2014).


\bibitem{Iachello2000}
        F. Iachello, Phys. Rev. Lett. {\bf 85}, 3580 (2000).

\bibitem{Caprio2007}
        M. A. Caprio, and F. Iachello, Nucl. Phys. A {\bf 781}, 26 (2007).

\bibitem{Wybourne1974}
        B. G. Wybourne, {\it Classical Groups for Physicists} (Wiley, New York, 1974).

\bibitem{Feinsilver1996}
        P. Feinsilver, Acta Appl. Math. {\bf 43}, 289 (1996).

\bibitem{Bonatsos2008}
        D. Bonatsos, E. A. McCutchan, and R. F. Casten,
        Phys. Rev. Lett. {\bf 101}, 022501 (2008).

\bibitem{Ui1970}
        H. Ui, Prog. Theor. Phys. {\bf 44}, 153 (1970).

\bibitem{Rowe2004IV}
        D. J. Rowe, Nucl. Phys. A {\bf 735}, 372 (2004).

\bibitem{Rowe2005}
        D. J. Rowe and P. S. Turner, Nucl. Phys. A {\bf 753}, 94 (2005).

\bibitem{Rowe2005II}
        D. J. Rowe, J. Phys. A {\bf 38}, 10181 (2005).

\bibitem{Rowe2009}
        D. J. Rowe, T. A. Welsh, and M. A. Caprio, Phys. Rev. C {\bf 79}, 054304 (2009).

\bibitem{Pan1998}
        F. Pan and J. P. Draayer, Nucl. Phys. A {\bf 636}, 156 (1998).


\bibitem{Pan2002}
        F. Pan, X. Zhang, and J. P. Draayer, J. Phys. A {\bf 35}, 7173 (2002).


\bibitem{Warner1983}
        D. D. Warner, and R. F. Casten,
        Phys. Rev. C {\bf  28}, 1798 (1983).

\bibitem{Iachello2004}
        F. Iachello and N. V. Zamfir,
        Phys. Rev. Lett. {\bf  92}, 212501 (2004).

\bibitem{Iachello1998}
        F. Iachello, N. V. Zamfir, and R. F. Casten
        Phys. Rev. Lett. {\bf  81}, 1191 (1998).



\bibitem{Edmondsbook}
        A. R. Edmonds, {\it Angular Momentum in Quantum Mechanics} (Princeton University Press, Princeton, 1957).




\bibitem{Arias2003}
        J. M. Arias, J. Dukelsky, and J. E. Garc\`{i}a-Ramos,
        Phys. Rev. Lett. {\bf  91}, 162502 (2003).


\bibitem{PZD2005}
        F. Pan, Y. Zhang, and J. P. Draayer,
        J. Phys. G {\bf 31}, 1039 (2005).

\bibitem{Dusuel2005}
        S. Dusuel, J. Vidal, J. M. Arias, J. Dukelsky, and J. E. Garc\`{i}a-Ramos,
        Phys. Rev. C {\bf 72}, 011301(R) (2005).

\bibitem{ZL2002}
        D. L. Zhang and Y. X. Liu,
        Phys. Rev. C {\bf 65}, 057301 (2002).

\bibitem{Casten2000}
        R. F. Casten and N. V. Zamfir Phys. Rev. Lett. {\bf 85}, 3584 (2000).

\bibitem{Frank2001}
        A. Frank, C. E. Alonso, and J. M. Arias, Phys. Rev. C {\bf 65}, 014301 (2001).


\bibitem{Zamfir2002}
        N. V. Zamfir, {\it et al.}, Phys. Rev. C {\bf 65}, 044325 (2002).

\bibitem{Clark2004}
        R. M. Clark, {\it et al.}, Phys. Rev. C {\bf 69}, 064322 (2004).

\bibitem{Mihai2007}
        C. Mihai, {\it et al.}, Phys. Rev. C {\bf 75}, 044302 (2007).

\bibitem{Zhang2003}
        J. F. Zhang, {\it et al.}, Chin. Phys. Lett. {\bf 20}, 1231 (2003).

\bibitem{Conquard2009}
        L. Coquard, {\it et al.}, Phys. Rev. C {\bf 80}, 061304(R) (2009)

\bibitem{Blachot1997}J. Blachot, Nucl. Data Sheets {\bf 81}, 599 (1997).

\bibitem{Sonzogni2004}A. A. Sonzogni, Nucl. Data Sheets {\bf 103}, 1 (2004).

\bibitem{singh1996}B. Singh, Nucl. Data Sheets {\bf 78}, 395 (1996).

\bibitem{Blachot2002}J. Blachot, Nucl. Data Sheets {\bf 97}, 593 (2002).

\bibitem{Bonatsos2006}
        D. Bonatsos, D. Lenis, N. Pietralla, and P. A. Terziev
        and A. Frank, Phys. Rev. C {\bf 74}, 044306 (2006).

\bibitem{AAVGDF2003}
        J. M. Arias, C. E. Alonso, A. Vitturi, J. E. Garc\`{i}a-Ramos, J. Dukelsky,
        and A. Frank, Phys. Rev. C {\bf 68}, 041302(R) (2003).


\bibitem{Ramos2008}
        J. E. Garc\`{i}a-Ramos
        and J. M. Arias, Phys. Rev. C {\bf 77}, 054307 (2008).

\bibitem{Zhang2011}
        Y. Zhang, F. Z. Hou, and Y. X. Liu, and Y. Sun, Sci. China Phys. Mech. Astron. {\bf 54}, 227 (2011).




\end{thebibliography}
\end{document}